\documentclass[pdflatex,sn-mathphys-num]{sn-jnl}

\usepackage{graphicx}%
\usepackage{multirow}%
\usepackage{amsmath,amssymb,amsfonts}%
\usepackage{amsthm}%
\usepackage{mathrsfs}%
\usepackage[title]{appendix}%
\usepackage{xcolor}%
\usepackage{textcomp}%
\usepackage{manyfoot}%
\usepackage{booktabs}%
\usepackage{algorithm}%
\usepackage{algorithmicx}%
\usepackage{algpseudocode}%
\usepackage{listings}%
\usepackage{natbib}%

\usepackage{booktabs}
\usepackage{array}
\usepackage{siunitx}
\usepackage{tablefootnote}
\usepackage{tabularx}

\theoremstyle{thmstyleone}%
\theoremstyle{thmstyletwo}%

\theoremstyle{thmstylethree}%

\begin{document}

\title[Article Title]{Robust stabilization of high-frequency magnetic droplets in W-CoFeB-MgO nanoconstriction spin Hall nano-oscillators}


\author[1]{\fnm{Hind} \sur{Prakash}}
\author[1]{\fnm{Arunima} \sur{TM}}
\author[2]{\fnm{Roman} \sur{Khymyn}}

\author*[1]{\fnm{Himanshu} \sur{Fulara}}\email{himanshu.fulara@ph.iitr.ac.in}


\affil[1]{\orgdiv{Department of Physics}, \orgname{Indian Institute of Technology Roorkee}, \orgaddress{\city{Roorkee}, \postcode{247667}, \state{Uttarakhand}, \country{India}}}

\affil[2]{\orgdiv{Physics Department}, \orgname{University of Gothenburg}, \orgaddress{ \postcode{412 96}, \state{Gothenburg}, \country{Sweden}}}




\abstract{Magnetic droplets are highly nonlinear spin-wave solitons that can be excited in nanoscale spintronic devices with strong perpendicular magnetic anisotropy. Although extensively studied in nanocontact-based spin-torque oscillators, their stabilization in pure spin current-driven devices such as spin Hall nano-oscillators (SHNOs) has remained elusive. Here, we micromagnetically demonstrate the robust stabilization of non-propagating high-frequency droplets in CMOS-compatible W/CoFeB/MgO nanoconstriction SHNOs under oblique magnetic fields. While the constriction geometry gives rise to noncircular droplet shapes in an inhomogeneous effective field landscape, stable droplet modes exhibiting complete core magnetization reversal and pronounced hysteresis are observed. At lower current densities, droplets display breathing oscillations with periodic expansion and contraction, whereas higher drive currents lead to drift, deformation, and the emergence of sidebands around the fundamental frequency. Tuning the strength and orientation of the applied magnetic field alters the effective field landscape, allowing droplets to escape confinement and propagate over distances exceeding 2$\mu$m.}




\maketitle

\section*{Introduction}\label{sec1}

Magnetic solitons~\cite{ivanov1989pl,kosevich1990pr}, particle-like localized spin configurations in magnetically ordered systems, have drawn considerable attention for their rich nonlinear dynamics and potential applications in next-generation microwave and neuromorphic technologies~\cite {torrejon2017neuromorphic,romera2018nature}. These excitations can be either topologically nontrivial, such as magnetic skyrmions~\cite{jiang2015sc,marrows2021apl}, or topologically trivial, such as magnetic droplets~\cite{hoefer2010prb,mohseni2013sc}. Magnetic droplet solitons are localized spin-wave (SW) excitations that arise in thin ferromagnetic films with strong perpendicular magnetic anisotropy (PMA) ~\cite{mohseni2013sc,macia2014ntn,iacocca2014prl,ahlberg2022ntc,Ahlberg2024,Kuchkin2025CommPhys}. They can be viewed as dissipative counterparts of the "magnon drops" originally predicted for ideal, lossless media ~\cite{ivanov1989pl,kosevich1990pr}. In ferromagnetic nanoscale devices, where intrinsic damping is inevitable, droplet solitons were first theoretically predicted~\cite{hoefer2010prb} and subsequently experimentally realized in nanocontact spin-torque nanooscillators (STNOs) ~\cite{mohseni2013sc,macia2014ntn,backes2015prl,burgos2018sr,chung2018prl,jiang2024apr}. In these devices, spin-transfer torque (STT) from a spin-polarized current drives large-amplitude, localized SW auto-oscillations in a ferromagnetic layer with strong PMA, enabling tunable GHz-frequency nanoscale microwave sources with high output power and compatibility with neuromorphic computing architectures~\cite{romera2018nature,Zahedinejad2019NatN}.

\begin{figure}
  \begin{center}
  \includegraphics[width=1\textwidth]{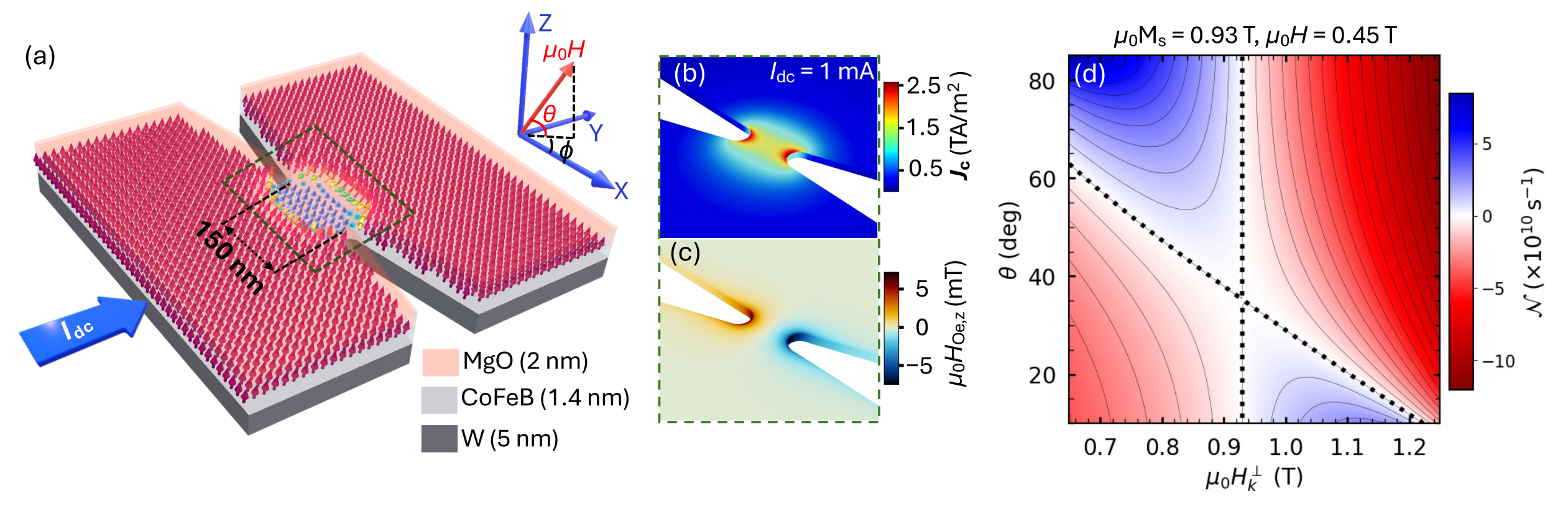}
    \caption{\textbf{SHNO device schematic, COMSOL simulations, and magnetodynamical nonlinearity.} \textbf{a,} Schematic of a 150 nm nanoconstriction SHNO device, illustrating magnetic droplet nucleation at the constriction region. COMSOL simulated \textbf{b,} Current density and \textbf{c,} Oersted field distributions for the highlighted area in the schematic. \textbf{d,} Color map of the analytically calculated nonlinearity coefficient ($\mathcal{N}$) for a ferromagnetic thin film (employing the methodology outlined in ref.~\cite{gerhart2007prb,slavin2008ieee}), shown as a function of PMA field and applied  OOP field orientation ($\mu_0 H = 0.45$ T), for $\mu_0M_\mathrm{s} = 0.93$ T. Dotted black lines indicate $\mathcal{N}$ = 0.}
    \label{fig:1}
    \end{center}
\end{figure}

A complementary class of spintronic oscillators, known as spin Hall nano-oscillators (SHNOs)~\cite{demidov2012ntm,Chen2016procieee,demidov2017phyreports,arunima2025jpd}, utilizes current-induced spin-orbit torque (SOT)~\cite{brataas2014natnano,Fulara2019SciAdv} to drive magnetization dynamics, offering a highly energy-efficient route for the generation of droplet solitons in flexible nonmagnetic-ferromagnetic heterostructures~\cite{chen2020pra,Klause2022apl,Ovcharov2024apl}. The first experimental signatures of droplet excitations in such devices were reported by Divinskiy et al.\cite{Divinskiy2017prb} in nanoconstriction-based SHNOs~\cite{demidov2014apl,Awad2016NatPhys,arunima2025apl}. Compared with conventional STNOs, nanoconstriction SHNOs offer three key advantages: (i) direct optical access to the active region~\cite{demidov2014apl,Muralidhar2022optothermal}, allowing spatially resolved mapping of SW intensity, (ii) scalable mutual synchronization of multiple oscillators through a single drive current~\cite{Awad2016NatPhys,Kumar2023robust}, and (ii) compatibility with hybrid device architectures in which an in-plane current excites auto-oscillations, while a magnetic tunnel junction (MTJ) provides efficient microwave readout~\cite{Tarequzzaman2019CommPhys,cai2023edl}. In addition, SHNOs exhibit rich nonlinear spin-dynamics under both in-plane (IP) and out-of-plane (OOP) magnetic fields\cite{demidov2014apl,demidov2017phyreports,Dvornik2018PRA}, which include edge-localized modes~\cite{Dvornik2018PRA,hache2025nl,arunima2025jpd}, spin-wave bullets~\cite{demidov2014apl,Rajabali2023PRA}, and propagating spin-wave modes~\cite{Fulara2019SciAdv,kumar2025NatPhy}. Among various SHNOs material systems, W/CoFeB/MgO nanoconstriction SHNOs (Fig.~\ref{fig:1}(a)) stand out due to their CMOS-compatible fabrication~\cite{Zahedinejad2018APL,Fulara2019SciAdv,Behera2022energy} and the voltage-controlled tuning of oscillator synchronization via gate modulation~\cite{Fulara2020natcomm,Zahedinejad2022natmat}. Divinskiy et al.~\cite{Divinskiy2017prb} observed propagating magnetic droplets in nanoconstriction SHNOs, driven by spatially extended SOTs. However, the strongly nonuniform effective field landscape in these geometries typically inhibits the stabilization of non-propagating droplets with a well-defined frequency, in contrast to the behavior observed in STNOs. This limitation raises a fundamental question that motivates the present work: \textit{can non-propagating magnetic droplets be stabilized within such a strongly inhomogeneous field landscape?}

Here, we address this problem by combining nonlinear auto-oscillator theory with micromagnetic simulations to elucidate the nucleation mechanism and dynamic behavior of both stationary and propagating droplets in CMOS-compatible W/CoFeB/MgO nanoconstriction SHNOs with strong PMA. We demonstrate the formation of robust, non-propagating high-frequency magnetic droplets that remain strongly localized well below the ferromagnetic resonance (FMR) frequency. At the threshold current, these droplets emerge as breathing modes, which evolve at higher drive currents into drifting, shape-distorted droplets accompanied by spectral sidebands around the fundamental frequency. Moreover, while stronger OOP fields promote droplet localization, reducing the field strength toward the in-plane direction facilitates the transition into propagating droplets. From an application perspective, both confined and propagating magnetic droplets offer distinct functionalities. Confined droplets remain localized beneath the nanoconstriction and can mediate synchronization among neighboring SHNOs, providing a new route toward enhanced spectral coherence and microwave power generation. In contrast, propagating droplets can transport magnetic information over micrometer distances, making them attractive for magnonic interconnects and wave-based information processing. The ability to controllably switch between localized and propagating droplet regimes within a CMOS-compatible SHNO, together with their strong nonlinearity, hysteretic short-term memory, and reconfigurable dynamics, makes magnetic droplets promising building blocks for next-generation spintronic, magnonic, reservoir computing, and neuromorphic information processing technologies.

\section*{Results}\label{sec2}

\subsection*{Device schematic and magnetodynamic nonlinearity.}\label{subsec1}

Figure~\ref{fig:1}(a) shows the schematic of the SHNO device along with an illustration of a nucleated droplet within the constriction geometry, where arrows indicate the local magnetization direction. The constriction has a nominal width of 150 nm, an opening angle of 22$^\circ$, and a curvature radius of 50 nm. Figures~\ref{fig:1}(b) and (c) present the simulated current density and Oersted field distributions within the bilayer for a drive current, $I_{\mathrm{dc}}$ = 1 mA, focused on the highlighted region in the schematic (detailed in the
Methods section).

In a nonlinear auto-oscillator~\cite{slavin2005ieeem,slavin2008ieee}, the magnetization dynamics is governed by the nonlinearity coefficient, $\mathcal{N}$, which quantifies the nonlinear frequency shift of a magnetic nano-oscillator. A negative $\mathcal{N}$ corresponds to frequency redshifting with increasing oscillation amplitude, driving the self-localization of SW auto-oscillations and the formation of solitonic modes such as SW bullets~\cite{slavin2005prl,Mazraati2018pra} (in in-plane magnetized films) and magnetic droplets~\cite{mohseni2013sc,mohseni2018prb} (in perpendicularly magnetized films). In contrast, a large positive $\mathcal{N}$ results in frequency blueshifting above the FMR frequency, favoring SW propagation. The sign and magnitude of $\mathcal{N}$ are primarily determined by material parameters, such as saturation magnetization ($M_\mathrm{s}$), PMA field ($H_{\text{k}}^{\perp}$), and by the strength and orientation of the applied magnetic field. Although moderate PMA enables positive nonlinearity to be reached at comparatively low OOP fields, further increasing PMA beyond the level that fully compensates for shape anisotropy (i.e., when the equilibrium magnetization direction switches from in-plane to fully perpendicular) again results in a strongly negative $\mathcal{N}$~\cite{Fulara2019SciAdv}. To analytically determine the conditions favoring stable droplet formation in perpendicularly magnetized films ($H_{\text{k}}^{\perp} > M_\mathrm{s}$), we adopt the established framework detailed in~\cite{gerhart2007prb,slavin2008ieee}. Within this formalism, $\mathcal{N}$ is defined from the derivative of the FMR frequency $\omega_0$ with respect to $M_\mathrm{s}$ as~\cite{gerhart2007prb}  

\begin{equation}
\mathcal{N} \;\equiv\; -\,2M_\mathrm{s}\,\frac{\partial \omega_0}{\partial M_\mathrm{s}}\, .
\end{equation}

The FMR frequency is expressed as 

\begin{equation}
\omega_0^2 \;=\; \omega_\mathrm{H}\Bigl(\omega_\mathrm{H} + (\omega_\mathrm{M} - \omega_\mathrm{k})\cos^2\theta_{\mathrm{int}}\Bigr),
\end{equation}

where $\omega_\mathrm{H}=\gamma \mu_0 H_{\mathrm{int}}$ is the Zeeman frequency associated with the internal field $H_{\mathrm{int}}$, $\gamma$ is the gyromagnetic ratio, $\mu_0$ is the vacuum permeability, and $\theta_{\mathrm{int}}$ is the equilibrium magnetization angle with respect to the film normal. Here, $\omega_\mathrm{M}=\gamma \mu_0 M_\mathrm{s}$ denotes the demagnetizing contribution, and $\omega_\mathrm{k}=\gamma \mu_0 H_\text{k}^{\perp}$ represents the PMA contribution with $H_\text{k}^{\perp}=2K_\mathrm{u}/(\mu_0 M_\mathrm{s})$.

Substituting these terms into the derivative, the resulting expression for $\mathcal{N}$ in the presence of PMA:

\begin{equation}
\mathcal{N} = \frac{\omega_\mathrm{H}(\omega_\mathrm{M} - \omega_\mathrm{k})}{\omega_0}
\left(\frac{3\omega_\mathrm{H}^2\sin^2\theta_{\mathrm{int}}}{\omega_0^2} - 1\right).
\end{equation}

Figure~\ref{fig:1}(d) shows the calculated $\mathcal{N}$ as a function of PMA field strength and applied field ($\mu_0 H = 0.45$ T) orientation for $\mu_0 M_\mathrm{s} = 0.93$ T, with the boundary $\mathcal{N} = 0$ indicated by dotted lines. As shown in Figure~\ref{fig:1}(d), $\mathcal{N}$ becomes strongly negative (red regions) for large PMA fields and near perpendicular field orientations, driving SW self-localization and potentially enabling stable droplet nucleation.

To complement this analysis, we further study W/CoFeB/MgO-based nano-oscillators through micromagnetic simulations performed with MuMax3~\cite{vansteenkiste2014aip} by numerically solving the Landau–Lifshitz–Gilbert–Slonczewski (LLGS) equation for the CoFeB layer. Material parameters and details are provided in Method section.

\begin{figure}
  \begin{center}
  \includegraphics[width=1\textwidth]{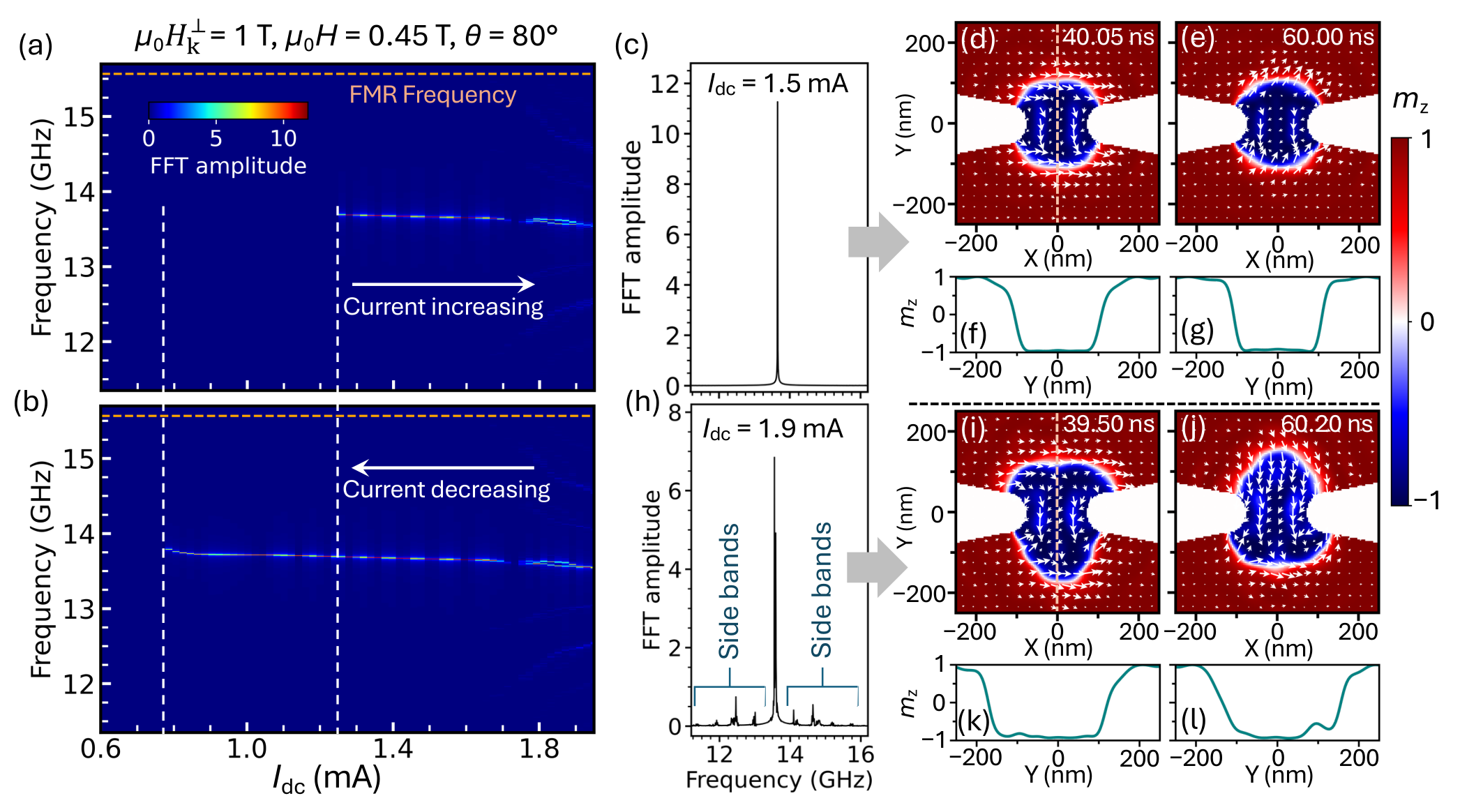}
    \caption{\textbf{Magnetic droplet dynamics in a 150 nm W/CoFeB/MgO nanoconstriction SHNO:} Current-sweep spectra for \textbf{a,} increasing and \textbf{b,} decreasing current exhibit pronounced hysteresis behaviour, evidencing droplet nucleation; the orange dashed line marks the FMR frequency. \textbf{c,} FFT spectrum at a fixed $I_{\mathrm{dc}} = 1.5$ mA with \textbf{d,e,} corresponding magnetization snapshots at two time instances and \textbf{f,g,} $m_\mathrm{z}$ line profiles along the $y$-axis through the constriction (orange dashed line), demonstrating complete magnetization reversal. \textbf{h,} FFT spectrum at $I_{\mathrm{dc}} = 1.9$ mA, highlighting droplet sidebands, with \textbf{i,j,} corresponding magnetization snapshots and \textbf{k,l,} $m_\mathrm{z}$ line profiles along the same axis, illustrating droplet asymmetry and drift.}
    \label{fig:2}
    \end{center}
\end{figure}

\subsection*{High-frequency non-propagating magnetic droplets.}\label{subsec1}

Since a large negative nonlinearity is essential for droplet stability~\cite{hoefer2010prb}, we first investigate the auto-oscillation dynamics of W/CoFeB/MgO SHNOs under an applied field ($\mu_0 H = 0.45$ T) oriented at $80^{\circ}$, for a PMA field ($\mu_0H_{\text{k}}^{\perp} = 1$T) near the compensation point (negative $M_{\mathrm{eff}}$), where $\mathcal{N}$ is strongly negative [Fig.~\ref{fig:1}(d)]. Figure~\ref{fig:2}(a) shows the micromagnetically simulated frequency spectrum during an increasing current sweep, with each step run for 150 ns after discarding the initial 40 ns to remove the transient effects. Auto-oscillations appear at a threshold current of 1.24 mA, followed by a gradual redshift as the drive current increases. In the corresponding decreasing current sweep [Fig.~\ref{fig:2}(b)], auto-oscillations persist down to 0.78 mA, revealing pronounced hysteresis~\cite{macia2014ntn}, a hallmark of droplet nucleation. The oscillation frequency lies well below the FMR (orange dashed line), confirming spatial localization within the constriction. The hysteresis indicates that the droplet state can be sustained at currents lower than those required for nucleation, implying an energy barrier between the droplet and the ferromagnetic ground state. Note that the hysteresis is absent in the propagating SW mode. Up to 1.65 mA, the frequency spectrum exhibits a single peak [Fig.~\ref{fig:2}(c) for $I_{\mathrm{dc}} = 1.5$ mA]. The corresponding magnetization snapshots and $m_\mathrm{z}$ line profiles [Figs.~\ref{fig:2}(d–g)] show complete magnetization reversal within the constriction, consistent with a stable droplet core. At the boundary, the magnetization oscillates nearly in phase, periodically expanding and contracting under STT~\cite{mohseni2013sc,Divinskiy2017prb}, characteristic of the droplet breathing mode. For $I_{\mathrm{dc}} > 1.65$ mA, sidebands emerge symmetrically around the fundamental frequency, indicating excitation of additional modes. At $I_{\mathrm{dc}} = 1.9$ mA [Fig.~\ref{fig:2}(h)], the sidebands become pronounced, and the overall FFT amplitude decreases as the energy is redistributed among these modes. The corresponding magnetization snapshots and $m_\mathrm{z}$ profiles [Figs.~\ref{fig:2}(i–l)] reveal droplet distortion and drift~\cite{mohseni2013sc,lendinez2015prb,liu2015prl,mohseni2020prb}. Overall, the formation of stable magnetic droplets in our simulations is thus established through four characteristic signatures: (i) oscillation frequency well below the FMR frequency, (ii) pronounced hysteresis in the frequency–current response, evidencing an energy barrier, (iii) complete magnetization reversal forming a localized droplet core, and (iv) the emergence of spectral sidebands associated with droplet drift. Finally, we investigate the robustness of the droplet stability under realistic conditions. Electro-thermal simulations show that Joule heating in the nanoconstriction region of the W/CoFeB/MgO SHNO produces only a modest temperature rise, insufficient to significantly alter the material properties, while room-temperature micromagnetic simulations confirm that the droplet remains stable in the presence of thermal fluctuations~\cite{Kuchkin2022prb}. We further assess the impact of material inhomogeneity by introducing structural grains with a 10\% Gaussian variation in PMA magnitude and a $2^{\circ}$ variation in the anisotropy axis. Although slight changes in the droplet shape and dynamics are observed, the localized droplet state remains stable, demonstrating its robustness against realistic thermal and material variations.

\subsection*{Phase diagrams of magnetic droplet dynamics.}\label{subsec1}

\begin{figure}
  \begin{center}
  \includegraphics[width=1\textwidth]{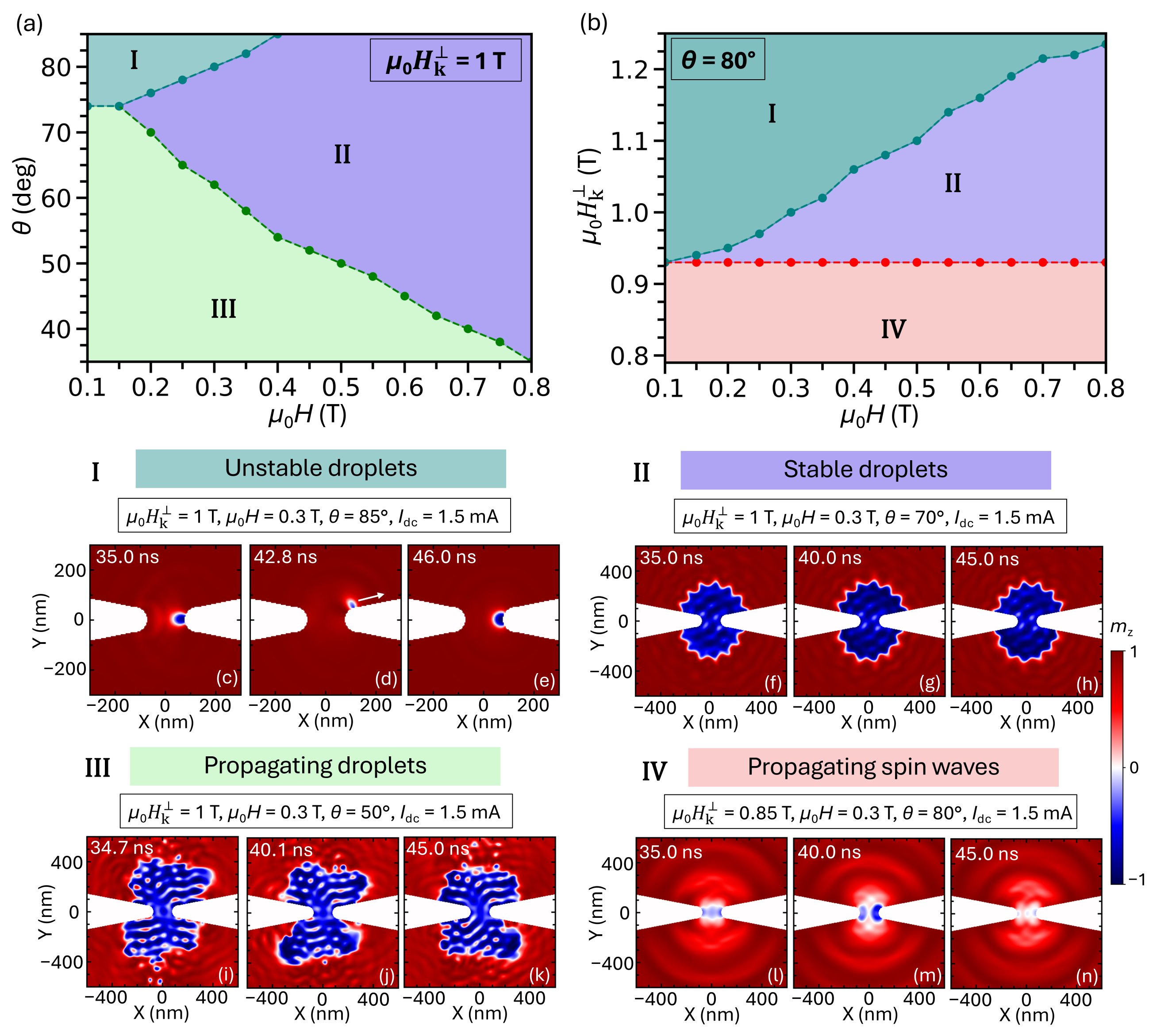}
    \caption{\textbf{Phase diagrams of auto-oscillation modes and representative magnetization dynamics.}  \textbf{a,} Phase diagram of auto-oscillations as a function of OOP field strength and field orientation at a fixed PMA of 1 T, revealing three regimes: unstable droplets (dark green), stable non-propagating droplets (purple), and propagating droplets (light green). \textbf{b,}  Phase diagram as a function of applied field and PMA at a fixed OOP angle of $80^{\circ}$, showing three regimes, including propagating SW auto-oscillations (light red). Representative magnetization snapshots of \textbf{c,d,e} Dynamically unstable droplets; \textbf{f,g,h,} a stable, non-propagating droplet confined to the constriction; \textbf{i,j,k,} cpropagating droplets breaking away from a stable core; and \textbf{l,m,n,} propagating spin waves.}
    \label{fig:3}
    \end{center}
\end{figure}

Next, we investigate magnetization dynamics under varying applied field strengths, field orientations, and PMA strengths, with a particular emphasis on the resulting droplet modes [Fig.~\ref{fig:3}(a-b)]. Figure \ref{fig:3}(a) shows the phase diagram as a function of applied field strength and OOP field orientation at a fixed PMA of 1 T (negative $M_{\mathrm{eff}}$), revealing three distinct regimes. The OOP angle range for the droplet formation is chosen from the nonlinearity map in Fig. \ref{fig:1}(d), ensuring operation within the large negative nonlinearity regime. In region $\textbf{I}$ (dark green), corresponding to near-perpendicular low fields, droplets nucleate at the edges of the constriction but remain dynamically unstable due to inhomogeneous field landscapes~\cite{iacocca2014prl,Divinskiy2017prb} [Fig.~\ref{fig:3}(c-e)]. This regime requires increasingly large OOP angles with increasing field strength; for example, at 0.4 T, the field angle must exceed $85^{\circ}$. The region $\textbf{II}$ (purple) hosts stable, non-propagating droplets localized within the constriction, characterized by a static core and a periodically oscillating boundary [Fig.~\ref{fig:3}(f-h)]. The size, shape, and boundary dynamics of the droplets evolve in response to the field strength, orientation, and current amplitude, resulting in breathing, drift, shape deformation, and perimeter-mode excitations. Stable droplets are favored at larger OOP fields over a wider angular range, with no stable states observed below 0.15 T. Additional simulations further demonstrate that droplet stability is preserved under realistic damping variations and in the presence of the expected interfacial Dzyaloshinskii--Moriya interaction (DMI) in W/CoFeB/MgO thin films, although finite DMI modifies the droplet shape ~\cite{Jaiswal2017apl}. Region $\textbf{III}$ (light green) marks the emergence of propagating droplets that break away from an expanded static core and move outward, typically occurring at lower fields with larger in-plane components. Figure \ref{fig:3}(b) presents a second phase diagram showing the dependence on magnetic field strength and PMA field at a fixed field orientation of 
$80^{\circ}$, highlighting the role of PMA. At large PMA field values, edge droplets nucleate ($\textbf{I}$, dark green); with increasing field strength, these droplets transition into stable droplets once the PMA exceeds a critical threshold of 0.93 T ($\textbf{II}$, purple). Below this value, the nonlinearity of the system favors SW propagation~\cite{mohseni2018prb,Fulara2019SciAdv} ($\textbf{IV}$, light red).

Representative magnetization snapshots illustrate these distinct regimes discussed above. Figures~\ref{fig:3}(c–e) show edge droplet nucleation at high OOP angles, where droplets attach to a constriction edge, propagate outward, annihilate, and reform, consistent with earlier reports~\cite{iacocca2014prl,Divinskiy2017prb}. Stable non-propagating droplets (region $\textbf{II}$) are shown in Figs.~\ref{fig:3}(f–h) at $\mu_0H = 0.3$ T, $\theta = 70^{\circ}$, and $I_{\mathrm{dc}} = 1.5$ mA. Under these conditions, the droplet exhibits a perimeter excitation mode (PEM), giving rise to a secondary spectral peak at half the fundamental frequency~\cite{Xiao2017prb}. The PEM is favored by stronger in-plane magnetic fields near the transition to the propagating-droplet regime and originates from exchange-driven restoring forces acting on deformations of the droplet boundary, resulting in standing-wave excitations with azimuthal periodicity. Propagating droplets (region $\textbf{III}$) are demonstrated in Figs.~\ref{fig:3}(i–k) at $\mu_0H = 0.3$ T, $\theta = 50^{\circ}$, and $I_{\mathrm{dc}} = 1.5$ mA, where a stripe-like static domain expands and emit droplets from its boundary. Finally, Figs.~\ref{fig:3}(l–n) illustrate propagating SWs observed at moderate PMA~\cite{Fulara2019SciAdv,kumar2025NatPhy}.

A key result of this study is the stabilization of non-propagating high-frequency magnetic droplets in nanoconstriction SHNOs over a broad range of operating conditions and constriction widths, despite the inherently inhomogeneous effective field landscape within the constriction region~\cite{Divinskiy2017prb,Ahlberg2024}. Their stability arises from the interplay of large negative nonlinearity, geometric pinning that forms a local potential well, and locally enhanced STT due to current crowding, which collectively suppresses drift and propagation. The applied magnetic field, through its IP and OOP components, plays a pivotal role in governing droplet stability and dynamics by altering the symmetry of the effective field and torque balance across the constriction. For small IP fields, the magnetization remains nearly perpendicular, but the weak asymmetry in the effective field leads to uneven precession amplitudes across the constriction. Given the highly nonuniform SOT generated by current crowding, this imbalance results in a net lateral force that pushes the droplet away from the potential minimum, yielding transient and unstable states. As the IP field increases to a moderate level, the field-induced canting of the magnetization introduces a compensating Zeeman torque that counteracts this imbalance, deepens the potential well, and stabilizes the droplet into a robust non-propagating state. This optimal IP-field regime coincides with a favorable OOP tilt—strong enough to maintain large negative nonlinearity and droplet self-localization, yet not so strong that the magnetization is nearly saturated out of plane. At larger IP fields, or equivalently when the effective OOP component becomes too weak, the PMA and nonlinear confinement diminish, flattening the potential landscape. In this regime, Zeeman energy outweighs the self-localizing nonlinear term, pushing the droplet along the spin-current direction and transforming it into a propagating soliton periodically emitted from the constriction. Similarly, very large OOP tilts suppress nonlinearity and destabilize confinement, preventing sustained droplet localization. Overall, the transition from unstable to confined to propagating droplet modes reflects a finely tuned interplay among SOT, Zeeman energy, geometric pinning, and nonlinear magnetic confinement, highlighting rich opportunities for reconfigurable droplet control in next-generation spintronic architectures.

\begin{figure}
  \begin{center}
  \includegraphics[width=0.6\textwidth]{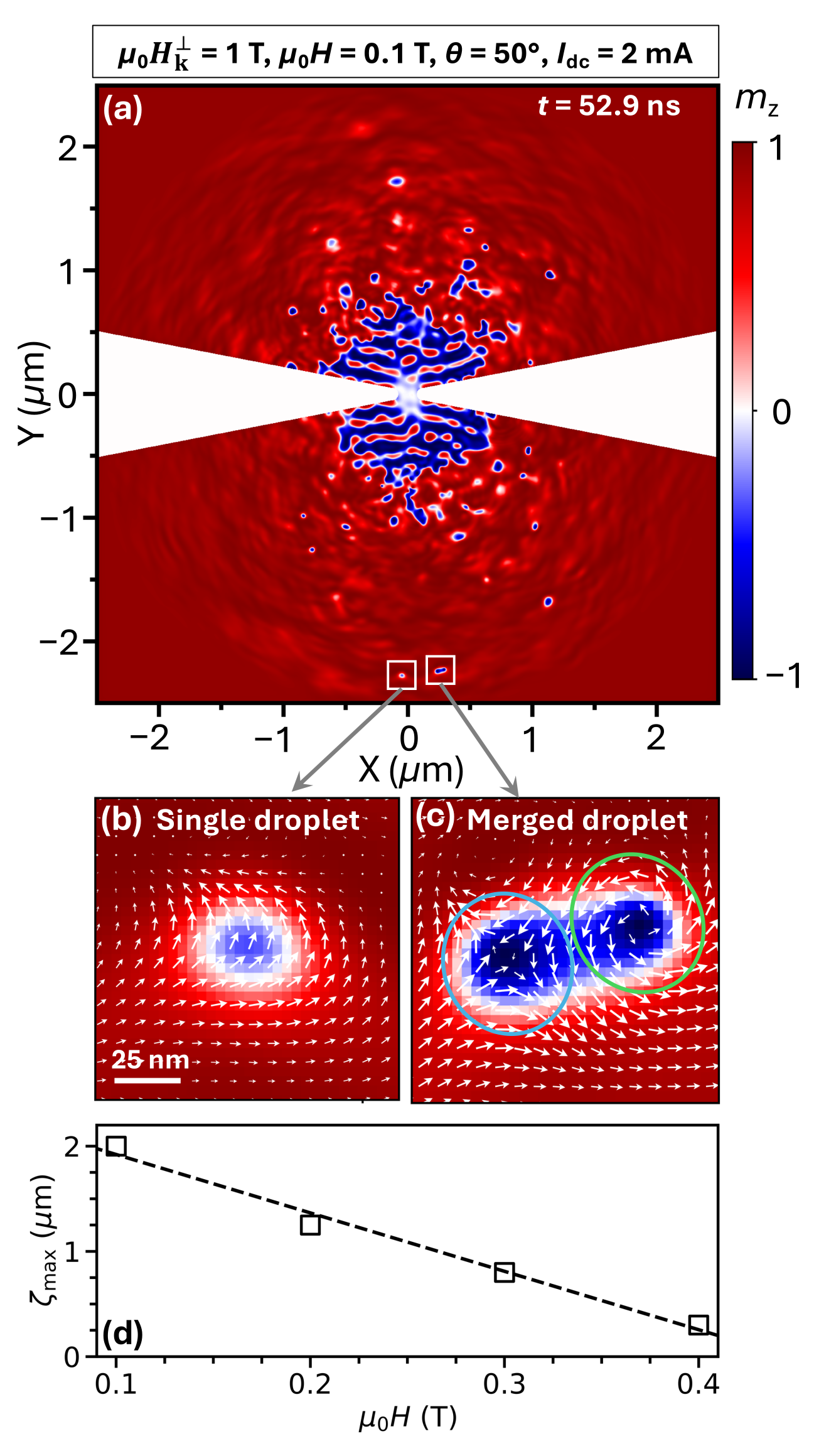}
    \caption{\textbf{Propagating magnetic droplets and their propagation length.}  \textbf{a,} Magnetization snapshot at $t=52.9$ ns showing droplet propagation in the nanoconstriction SHNO. The zoomed-in views in \textbf{b,} and \textbf{c,} highlight a single propagating droplet and a merged droplet state, respectively. \textbf{d,} Maximum droplet propagation length as a function of the applied magnetic field; the black dashed line represents a linear fit to the data points.}
    \label{fig:4}
    \end{center}
\end{figure}

\subsection*{Long range propagation of magnetic droplets.}\label{subsec1}

Finally, we investigate the dynamics of propagating droplets, specifically their type and maximum propagation length ($\zeta_\mathrm{max}$), within regime $\textbf{III}$ of the phase diagram in Figure \ref{fig:3}(a). To quantify the propagation length, we simulate a larger device area of $5~\mu\mathrm{m}\times5~\mu\mathrm{m}$ while maintaining the same cell size as in previous simulations, discretized into a $1280\times1280$ grid. Absorbing boundary conditions are applied to suppress reflections of spin waves generated by droplet annihilation at the device edges. Figure~\ref{fig:4}(a) shows a representative magnetization snapshot at $t=52.9$ ns for $\mu_0H_{\text{k}}^{\perp} = 1$ T, an applied field of 0.1 T at $50^{\circ}$, and a drive current of 2 mA. Under these optimal conditions, propagating droplets nucleate by breaking away from the boundary of an expanded stripe-like magnetic domain that extends approximately 0.5 $\mu$m from the constriction center. Once detached, the droplets are propelled outward by the SOT in the extended region. This propagation arises from the non-uniform spatial distribution of the drive current, which produces a gradient in current density on the scale of the droplet size, creating a spatially varying spin current and a boundary-dependent STT that drives droplet motion. Multiple propagating droplets can be generated concurrently, and mutual interactions occasionally lead to coalescence into a single merged droplet, consistent with previous observations by Maiden et al.~\cite{Maiden2014prb}. Zoomed-in views of single and merged droplets are shown in Figures~\ref{fig:4}(b) and \ref{fig:4}(c), respectively. Despite merging, the composite state retains a net topological charge of zero, identical to that of individual droplets, while exhibiting a hybrid spin texture comprising distinct skyrmion and antiskyrmion regions~\cite{Prakash2025apl,hoefer2010prb}. The continuous evolution of the local spin orientation causes periodic modulation of chirality, alternately bringing skyrmion and antiskyrmion regions closer together and farther apart along with the corresponding topological charge distribution. The dependence of the maximum propagation length $\zeta_\mathrm{max}$ on the applied magnetic field is summarized in Fig.~\ref{fig:4}(d) for a fixed \textit{dc} current. Increasing the field strength leads to a monotonic decrease in $\zeta_\mathrm{max}$, and beyond 0.5 T, propagating droplets cease to form, giving way to stable, non-propagating droplet modes confined within the constriction.

\newcounter{rowno}
\newcommand{\rownum}{\stepcounter{rowno}\therowno}
\newcolumntype{Y}{>{\centering\arraybackslash}X}

\begin{table}[htbp]
\centering
\color{black}
\caption{\textbf{Benchmarking of the present work with reported magnetic droplet studies.}}
\label{table:1}
\renewcommand{\arraystretch}{2}
\setlength{\tabcolsep}{1 pt}
\footnotesize
\setcounter{rowno}{0}
\begin{tabularx}{\textwidth}{@{} c Y Y Y Y Y Y Y @{}}
\toprule
\textbf{S. No.} & \textbf{Device} & \textbf{Study type} & \textbf{Material} & \textbf{Threshold current} & \textbf{Droplet stability} & \textbf{Frequency} & \textbf{Propagation\ length} \\
\midrule

\rownum & STNO~\cite{Hoefer2012prb} & Theory /Simulation & Ferromagnet (PMA) & -- & Stable & -- & 10 $\mu$m\textsuperscript{a} \\

\rownum & STNO~\cite{mohseni2013sc} & Experiment & Co/Ni & 4 mA  & Stable & 20--32 GHz & -- \\

\rownum & STNO~\cite{iacocca2014prl} & Simulation & Co/Ni & 4 mA & Stable & 10--15 GHz & -- \\

\rownum & STNO~\cite{macia2014ntn} & Experiment & Co/Ni & 25 mA & Stable & 16--25 GHz & -- \\

\rownum & STNO~\cite{Mohseni2020pra} & Simulation & Ferromagnet (PMA) & 4 mA\textsuperscript{b}  & Stable & 17--18 GHz & 0.6 $\mu$m\textsuperscript{c} \\

\rownum & STNO~\cite{chung2018prl} & Experiment &  Co/Ni & 12 mA & Stable & 5--18 GHz & --  \\

\rownum & STNO~\cite{ahlberg2022ntc} & Experiment /Simulation & Co/Ni & 4 mA & Stable & $<0.3$ GHz & -- \\

\rownum & STNO~\cite{Jiang2024ntc} & Experiment /Simulation & Co/Ni & 5 mA & Stable & $<3$ GHz & -- \\

\rownum & STNO~\cite{Ovcharov2024apl} & Theory /Simulation & Anti- ferromagnet & 48 mA & Stable & 205--520 GHz & -- \\

\rownum & SHNO~\cite{Divinskiy2017prb} & Experiment /Simulation & Co/Ni & \SI{3.5}{\milli\ampere} & Quasi-stable & 1--2.5 GHz & 1.5 $\mu$m\\

\rownum & SHNO~\cite{chen2020pra} & Experiment /Simulation & Co/Ni & 10 mA  & -- & $<2$ GHz & -- \\

\textbf{\rownum} & \textbf{SHNO (This work)} & \textbf{Simulation} & \textbf{CoFeB} & \textbf{1.24 mA}  & \textbf{Stable} & \textbf{$\sim$ 14 GHz}   & \textbf{2  \boldmath{$\mu$m}} \\

\bottomrule
\end{tabularx}
\vspace{2pt}
\raggedright\footnotesize
\textsuperscript{a} Propagation length for $\alpha=0.001$.\quad
\textsuperscript{b} Drive current pulse.  \quad
\textsuperscript{c}Propagation length for $\alpha=0.02$.
\end{table}

To put our results in context, Table~\ref{table:1} provides a quantitative comparison with previously reported theoretical and experimental studies on magnetic droplet solitons. The benchmarking includes threshold current, oscillation frequency, propagation lengths, and stability regimes. While magnetic droplets have been studied predominantly in Co/Ni-based STNOs, the present work demonstrates field-tunable droplet breathing, drift, and propagation in a CMOS-compatible W/CoFeB/MgO nanoconstriction SHNO. Note that among the limited studies on droplet dynamics in SHNOs, the present nanoconstriction-based SHNOs achieve droplet frequencies up to 14 GHz and propagation lengths exceeding 2 $\mu$m, surpassing previous SHNO demonstrations. This benchmarking highlights the unique capabilities and competitive performance of W/CoFeB/Mgo nanoconstriction SHNOs for future magnonic and spintronic applications.

\section*{Discussion}\label{sec3}

The capability to nucleate both stable non-propagating and propagating magnetic droplets in CMOS-compatible W/CoFeB/MgO nanoconstriction SHNOs offers distinct advantages that broaden the utility of these devices. Non-propagating droplets remain localized beneath the nanoconstriction, providing stable, phase-coherent, high-frequency auto-oscillations with minimal spectral broadening and robustness against perturbations. Unlike propagating droplets that rapidly leave the active region, these stable localized droplets can potentially mediate synchronization among neighboring SHNOs in chains or arrays, offering a new route toward enhanced microwave output power and spectral coherence. Given that existing synchronized SHNO networks remain limited by low output power, droplet-mediated synchronization could provide an effective mechanism for realizing high-performance microwave oscillators and large-scale spintronic oscillator networks. Beyond their fundamental importance for nonlinear SW dynamics, confined droplets combine strong nonlinearity with pronounced hysteresis, naturally providing intrinsic short-term memory and complex nonlinear state transformations. These properties make nanoconstriction SHNOs promising building blocks for reservoir computing and neuromorphic information processing, where nonlinear dynamics and memory are essential computational resources. Note that these droplet states remain stable under realistic operating conditions, including thermal fluctuations, material inhomogeneity, damping variations, and fabrication-induced parameter variations such as changes in constriction width, highlighting their experimental feasibility and robustness. On the other hand, propagating droplets convert the localized auto-oscillations into mobile solitonic wavepackets that can travel micrometer distances, effectively transforming the SHNO into a controllable droplet emitter. This opens opportunities for SW-based information transport and processing, including magnonic interconnects and logic elements in which information is carried by the dynamical state of the droplet. In particular, the ability to reversibly toggle between confined and propagating droplet regimes through subtle variations in field strength and orientation demonstrates a versatile platform for reconfigurable device functionality on demand. Thus, nanoconstriction SHNOs uniquely combine the benefits of high-frequency droplet stability, nonlinear memory, and controllable long-range droplet transport, making them attractive for scalable neuromorphic and magnonic computing technologies.

In summary, we have investigated the nucleation, stability, and dynamic control of magnetic droplet solitons in nanoconstriction SHNOs. The constriction geometry restricts the extended droplet boundary from being perfectly circular, yet the nucleated high-frequency droplets exhibit complete core reversal and noticeable hysteresis during current sweeps. The emergence of spectral sidebands at higher drive currents signifies droplet deformation and drift, marking the onset of dynamic instabilities. By systematically varying the applied field strength and orientation, we identify distinct dynamical regimes ranging from stable non-propagating droplets to unstable and propagating modes that detach from the expanded droplet boundary. The interplay between PMA and the effective magnetic field governs the excitation and stability of these modes. Below a critical PMA threshold, only propagating SWs are sustained, whereas sufficiently large PMA, combined with oblique magnetic fields, stabilizes robust, high-frequency droplets. In contrast, strong near-perpendicular or predominantly in-plane fields destabilize confinement, enabling droplet escape and long-range propagation over micrometer scales. These results highlight practical strategies for engineering controlled droplet stability and long-range droplet transport in CMOS-compatible nanoconstriction SHNOs.

\section*{Methods}\label{se4}

\subsection*{COMSOL simulations.}
The current density and Oersted field distributions were simulated in COMSOL Multiphysics~\cite{COMSOL} using the Magnetic Fields (mf) module under a stationary study, where a coil geometry analysis was defined to model the current carrying device. In the model, MgO is treated as a perfect insulator, and only the W (5 nm)/CoFeB (1.4 nm) bilayer is considered, with resistivities of 213 and 100 $\mu\Omega \cdot$cm, respectively~\cite{Fulara2019SciAdv}. Coupled electrothermal simulations were carried out to estimate Joule heating in the W (5 nm)/CoFeB (1.4 nm) nanoconstriction SHNO on a high-resistivity Si substrate (thermal conductivity: 130 W/m$\cdot$K). The simulations were performed at an ambient temperature of 293 K using the Electric Currents (ec), Heat Transfer in Solids (ht), and Electromagnetic Heating (emh) modules.

\subsection*{Micromagnetic simulations.}
The current density and Oersted field distributions obtained from COMSOL simulations are directly mapped into the MuMax3 simulation domain. The geometry is discretized into $3.9 \times 3.9 \times 1.4$ nm$^3$ cells, smaller than the exchange length, $\lambda_\mathrm{ex} = \sqrt{\tfrac{2A_\mathrm{ex}}{\mu_0 M_\mathrm{s}^2}} = 7.4$ nm, and the Bloch length, $\lambda_\mathrm{B} = \sqrt{\tfrac{A_\mathrm{ex}}{k_\mathrm{u}}} = 7.1$ nm. The charge current in the W layer is converted into a spin current using a spin Hall angle of $\theta_{\mathrm{SH}} = -0.41$~\cite{Fulara2019SciAdv}, with polarization lying in the film plane. The material parameters are chosen based on experimental reports~\cite{Fulara2019SciAdv,Fulara2020natcomm}: $\mu_0M_\mathrm{s} = 0.93$ T, exchange stiffness $A_\mathrm{ex} = 19 \times 10^{-12}$ J/m, Gilbert damping $\alpha = 0.023$, $\gamma / 2\pi = 29.9$ GHz/T, and PMA field $\mu_0H_{\text{k}}^{\perp} = 1$ T. The FMR frequency is determined by exciting the system with a magnetic field sinc pulse. To investigate the material inhomogeneity effect, the magnetic layer was divided into grains with an average size of 30 nm using a Voronoi tessellation approach. The PMA of each grain was independently assigned from a Gaussian distribution centered around $\mu_0 H_\text{k}^{\perp} = 1$ T, with a 10\% standard deviation in anisotropy magnitude and a 2$^\circ$ variation in the anisotropy axis orientation.

\backmatter

\bmhead{Supplementary information}
The additional supplementary data in support of the main text are available in the Supplementary Information section.

\section*{Declarations}

\backmatter

\bmhead{Acknowledgements}
This work is supported by the SRG/2023/002487 project funded by the Anusandhan National Research Foundation (ANRF), Department of Science \& Technology, Government of India. HP and ATM acknowledge the financial support from the Ministry of Education, Government of India. We also thankfully acknowledge the Institute Computer Center (ICC), IIT Roorkee, for providing a high-end computational facility to run simulations.
\bmhead{No Competing Interests:} 
The authors declare no competing financial or non-financial interests.
\bmhead{Data Availability Statement}
The datasets generated and/or analyzed during the current study are available in the Zenodo repository at \href{https://doi.org/10.5281/zenodo.18620611}{https://doi.org/10.5281/zenodo.18620611}

\bmhead{Code availability statement}
The underlying code for this study is not publicly available but may be made available to qualified researchers on reasonable request from the corresponding author. The open-source software package mumax$^3$ used for simulations is available free of charge at https://mumax.github.io/.
\bmhead{Author contributions}
H.F. initiated the project. H.P. performed the micromagnetic simulations and analyzed the data. A.TM. carried out analytic calculations of the nonlinearity coefficient. R.K. and H.F. contributed to the data analysis. All authors contributed to the interpretation of the results and co-wrote the manuscript. H.F. coordinated and supervised the project.





\end{document}